\documentstyle[preprint,aps,version2,epsfig]{revtex}
\input{epsf}
\begin{document}
\vspace{-80ex}
\begin{flushright}
\vspace{-2.0mm}
TUIMP-TH-97/80\\
hep-ph/9703408\\
\vspace{-3.0mm}
\end{flushright}

\draft
\begin{title} {\Large \bf Nonperturbative Explanation of the Enhancement 
\\Factors in QCD Sum Rule for the $~\rho~$ Meson}
\end{title}
\bigskip
\begin{center}
Shu-Quan Nie\\
{\it Department of Physics, Tsinghua University, Beijing 100084, China}\\
\bigskip
Yu-Ping Kuang~~~~Qing Wang~~~~ Yu-Ping Yi\\ 
{\it China Center of Advanced Science and Technology (World Laboratory),\\
      P. O. Box 8730, Beijing 100080, China\\and\\
Institute of Modern Physics, Tsinghua University, Beijing 100084,
         China\footnote{Mailing address.}}
\end{center} 
\vspace{0.6cm}
\centerline{ Abstract}
\vspace{-1.4cm}
\begin{abstract}
Taking the sum rule for the $~\rho~$ meson as an example, we study the 
possibility of explaining the phenomenological enhancement factors 
for certain terms in the vacuum expectation value of the operator product 
expansion in the QCD sum rule. 
We take a QCD motivated extended Nambu-Jona-Lasinio (ENJL) model as the low 
energy effective Lagrangian for QCD with which we calculate the 
nonperpturbative contributions to the vaccum condensate expansion to obtain 
the enhancement factors. Our result shows that such nonperturbative 
contrbutions can cause large enough enhancement factors which  can be 
consistent with the phenomenological values.
\end{abstract}
\vspace{-4.0mm}

\narrowtext

\newpage

\begin{center}
{\bf I.~INTRODUCTION}
\end{center}

QCD sum rule \cite{SVZ} is a useful semi-phenomenological approach to hadron 
physics in which nonperturbative vacuum condensate (VC) effects are taken
into account. The sum rule relates physical observables to certain
VC's through the vacuum expectation values (VEV's) of products
of current operators $~\langle Tj^A(x)j^B(0) \rangle~$ with the vacuum 
condensate expansion (VCE)

\begin{eqnarray}                       
i\int d^4x\,e^{iq \cdot x}\,\langle Tj^A(x)j^B(0)\rangle\,=\,F^{AB}(q)\sum_n\,
C_n(Q^2)\,\langle O_n\rangle~,
\end{eqnarray}
\noindent
where $~F^{AB}(q)~$ is a tensor factor characterizing the tensor
structure in the indices $~A~{\rm and}~B~$, $~Q^2\equiv -q^2~$, $~\langle O_n
\rangle~$'s are VC's of various dimensions. Usually, up to the VC's
of dimension-6 operators (four-quark operators) are taken into account in (1). 
In most applications of the QCD sum rule, the following approaches are taken:

\null\noindent
(i) as a basic assumption, the coefficients $~C_n~$'s, at least in the
first few terms, are assumed to be just the Wilson coefficients in the standard
operator product expansion (OPE) determined merely from perturbative QCD 
(PQCD), and nonperturbative effects all reside in the VC's $~\langle O_n
\rangle~$'s \cite{SVZ};

\null\noindent
(ii) $~\langle O_n\rangle~$'s are treated as free parameters 
determined by experimental inputs, and to reduce the number of free parameters
, people often make the simple {\it factorization} approximation to express 
the four-quark condensate (4QC) in terms of the two-quark condensate (2QC)
in the following way \cite{SVZ}.
\begin{eqnarray}                                                   
\langle{\bar \psi}\Gamma_1\psi{\bar \psi}\Gamma_2\psi\rangle\,\approx~ 
N^{-2}[~{\rm Tr} \Gamma_1~ {\rm Tr} \Gamma_2-{\rm Tr}(\Gamma_1\Gamma_2)~]
\langle\bar{\psi}\psi\rangle^2~,
\end{eqnarray}
\noindent
where $~\psi~$ is the three-flavor light quark field, the Lorentz structure of 
$~\Gamma_{1,2}~$ is $~1~$ for scalar (S), $~i\gamma_5~$ for pseudoscalar (P), 
$~\gamma_{\mu}~$ for vector (V), $~\gamma_5\gamma_{\mu}~$ for axial-vector 
(A), $~\sigma_{\mu\nu}~$ for tensor (T) (for color and flavor non-singlet, the 
color and flavor group generators $~\lambda_\alpha/2~$, $~t_i~$ should also be 
included ), and $~N=36~$ is a normalization constant \cite{SVZ}. 

However, a careful analysis of the QCD sum rule for the $~\rho~$ meson 
\cite{LNT} shows that the theoretical results are not consistent with the 
experiments and phenomenologically the demension-6 4QC term should be enhanced 
by a factor 
\begin{equation}                         
\kappa_{4Q}\approx 3-6.6
\end{equation}
for fitting the data \cite{LNT}. Studies of other processes with the
same approach such as the sum rules for baryons \cite{D} and pseudoscalar 
meson \cite{KPS} lead to similar conclusion that an enhancement factor 
around $~\kappa_{4Q}\approx 4~$ is needed for fitting the data. 

There is an alternative analysis from the finite energy sum rule which leads 
to similar but somewhat different results that the dimension-6 4QC term and 
the dimension-4 gluon condensate plus two-quark condensate terms
(Actually, the gluon condensate term dominates.) should be enhanced by 
factors
\begin{eqnarray}                                       
\kappa_{4Q}\approx 5-8,~~~~~~~~~~~~~~~~~~~~\kappa_G\approx 2-5\,,
\end {eqnarray} 
respectively \cite{BDLPR}.

 So far there is no successful theoretical explanation of the enhancement 
factors. In our previous paper \cite{WKYC}, we kept assumption 
(i) and looked at the possibility of explaining the enhancement factor 
$~\kappa_{4Q}~$ in (3) by taking account of the non-factorized parts of the 
4QC's, i.e. modifying approximation (ii). We took the QCD motivated Nambu-
Jona-Lasinio (NJL) model by Bijnens, Bruno, and de Rafael \cite{BBdR} as the 
low energy effective Lagrangian of QCD to calculate the non-factorized parts 
of the 4QC's by means of a nonperturbative method for the effective potential 
for local composite operators \cite{HK}, and the obtained
non-factorized parts of the 4QC's are of the next-to-the-leading order 
in the $~1/N_c~$ expansion. Our conclusion is that for reasonable values of 
the parameters in the model, the obtained  enhancement factor $~\kappa_{4Q}~$ 
is only a few percent larger than unity, so that it is too small to account 
for (3). The next reasonable trial is to consider certain modification
of assumption (i) since the $~\rho~$ meson lies below the chiral symmetry 
breaking scale wherein nonperturbative effects are important. Although an 
argument considering nonperturbative effects from instantons in the dilute gas 
approximation has been given in Ref. \cite{SVZ} expecting that assumption (i) 
may work up to about ten terms in (1), the analyses in Refs. \cite{LNT}
\cite{BDLPR} show that the real situation is not so optimistic. In this paper, 
inspired by a recent paper by Yamawaki and Zakharov \cite{YZ}, we study the 
possibility of explaining the enhancement factors by taking into account 
certain nonperturbative contributions to the VCE in which nonperturbative 
effects not only reside in the condensates but also affect the coefficients in 
the VCE so that assumption (i) is modified. We take the whole extended Nambu-
Jona-Lasinio (ENJL) Lagrangian in Ref. \cite{BBdR} as the low energy effective 
Lagrangian of the quark and gluon system, and calculate the nonperturbative 
contributions to the VCE from the NJL four-fermion interactions to the 
precision of the leading order of $~1/N_c~$ expansion. Our results show that, 
taking only four-fermion interactions of the scalar and pseudoscalar types 
\cite{BBdR}, such modification mainly affects the 4QC terms in the VCE by 
causing additional 4QC's with different tensor structures which leads to a 
fairly large contribution to $~\kappa_{4Q}~$. For reasonable parameters in the 
model, it is really possible to obtain a large enough $~\kappa_{4Q}~$ 
consistent with (3). We shall also show that, when including certain vector 
and axial-vector four-fermion interactions in the ENJL Lagrangian
\cite{BBdR}, it is also possible to explain the enhancement factors (4). This 
means that for low energy processes, QCD sum rule may give better results if 
the original assumption (i) is modified by taking account of more 
nonperturbative QCD effects.

This paper is organized as follows. Sec.II is the theoretical aspect
of this study. In Sec.III we present the calculations of the nonperturbative 
contributions to the VCE up to the 4QC terms from the ENJL model and
the explanation of the enhancement factors. Conclusions are 
given in Sec.IV. \\

\begin{center}
{\bf II.~THE THEORETICAL STRATEGY }
\end{center}

To study nonperturbative contributions, it is convenient to take a low energy 
effective Lagrangian of QCD including nonperturbative effects. So far
there is no successful low energy effective Lagrangian derived from the
first principles of QCD. The QCD motivated ENJL model by Bijnens, Bruno
and de Rafael \cite{BBdR} contains reasonable QCD ingredient including
chiral symmetry breaking, and can lead to rather successful phenomenological 
results \cite{BBdR}. So we take this model as the low energy effective 
Lagrangian for our calculations. The Lagrangian in the ENJL model in Ref. 
\cite{BBdR} is
\begin{eqnarray}            
{\cal L}_{QCD}\,=\,{\cal L}^{\Lambda_\chi}_{QCD}\,+\,{\cal L}^{SP}_{NJL}~
+\,{\cal L}^{VA}_{NJL}~,
\end{eqnarray}
\noindent
where $~\Lambda_\chi~$ is a momentum cut-off below which the model serves as
the low energy effective Lagrangian for QCD, $~{\cal L}^{\Lambda_\chi}_
{QCD}~$ is of the same form of the $~QCD~$ Lagrangian for momentum below 
$~\Lambda_\chi~$, and
\begin{equation}        
\begin{array}{ll}
{\cal L}^{SP}_{NJL}\,=\,\displaystyle{{\frac {8\pi^2G_S}{N_c\Lambda^2_\chi}}
\sum_{ab}({\bar \psi}^a_R\psi^b_L)({\bar \psi}^b_L\psi^a_R)}~\\~\;\;\;
\;\;\;\;\;\;\,=\,\displaystyle{\frac{2\pi^2G_S}{3N_c\Lambda^2_\chi}\left[({\bar
\psi}\psi)^2\,+\,({\bar \psi}i\gamma_5\psi)^2 \right]}~\\
~\;\;\;\;\;\;\;\;\;\;\;\;\;\,
+\,\displaystyle{{\frac {4\pi^2 G_S}{N_c\Lambda^2_\chi}}
\sum_{i=1}^{8}\left[({\bar \psi}t_i\psi)^2\,+\,({\bar \psi}
i\gamma_5t_i\psi)^2\right]}~,
\end{array}
\end{equation}

\begin{equation}       
\begin{array}{ll}
{\cal L}^{VA}_{NJL}\,=\,-\,\displaystyle{{\frac {8\pi^2G_V}{N_c\Lambda^2_\chi}}
\sum_{ab}\left[({\bar \psi}^a_L\gamma^{\mu}\psi^b_L)({\bar \psi}^
b_L\gamma_{\mu}\psi^a_L)\,+\,({\bar \psi}^a_R\gamma^{\mu}\psi^b_R)
({\bar \psi}^b_R\gamma_{\mu}\psi^a_R)\right]}~\\~\;\;\;\;\;\;\;\;\;\,=\,-\,
\displaystyle{\frac{4\pi^2G_V}{3N_c\Lambda^2_\chi}}\left[({\bar \psi}
\gamma^{\mu}\psi)^2\,+\,({\bar \psi}\gamma_5\gamma^{\mu}\psi)^2
\right]~\\~\;\;\;\;\;\;\;\;\;\;\;\;\;\,-\,
\displaystyle{{\frac {8\pi^2 G_V}{N_c\Lambda^2_\chi}}\sum_{i=1}^{8}\left[
({\bar \psi}\gamma^{\mu}t_i\psi)^2\,+\,
({\bar \psi}\gamma_5\gamma^{\mu}t_i\psi)^2\right]}~,
\end{array}
\end{equation}
\noindent
in which $~G_S~$ and $~G_V~$ are two coupling constants treated as free 
parameters, and the flavor group generator $~t_i~$ is normalized as 
$~{\rm tr}(t_it_j)=\frac{1}{2}\delta_{ij}~$. It is argued in Ref. \cite{BBdR} 
that the NJL type Lagrangian $~{\cal L}^{SP}_{NJL}\,+\,{\cal L}^{VA}_{NJL}~$ 
may be understood as coming from integrating out the high momentum modes of 
quarks and gluons in the fundamental theory of QCD (However, $G_S$ and $G_V$ 
are taken as two independent free parameters here.) and is regarded as the 
part of $~{\cal L}_{QCD}~$ responsible for chiral symmetry breaking, and 
$~{\cal L}^{\Lambda_\chi}_{QCD}~$ provides perturbative corrections to the 
broken chiral symmetry state. This supports the idea of the chiral quark model 
\cite{MG}. 

Our strategy is that we apply the Lagrangian (5)-(7) to energy scale lower 
than $\Lambda_\chi$, and take the ordinary PQCD approach for the energy
range higher than $\Lambda_\chi$ where nonperturba-tive effects are 
not important. Since the structure of ${\cal L}^{\Lambda_\chi}_{QCD}$ is the 
same as the original QCD Lagrangian, {\it the pure perturbation results in
the present appraoch are the same as those from the ordinary PQCD}. 
{\it The NJL Lagrangian ${\cal L}^{SP}_{NJL}+{\cal L}^{VA}_{NJL}$ not only
gives rise to chiral symmetry breaking, but also give nontrivial
contributions to the VCE which is what we are going to calculate}.

There are three independent free parameters $~G_S~$,$~~G_V~$, and
$~\Lambda_\chi~$ in the ENJL model in Ref. \cite{BBdR}. They are related to
the dynamical quark mass $~M_Q~$, the quark axial-vector coupling
constant $~g_A~$, and the vector meson mass $~M_\rho~$ by the following
relations\cite{BBdR}

\begin{eqnarray}               
1/G_S\,=\,(M_Q/\Lambda_\chi)^2\Gamma(-1,\,(M_Q/\Lambda_\chi)^2)
(1\,+\,\gamma_{-1})\,,
\end{eqnarray}

\begin{eqnarray}               
\displaystyle{g_A\,=\,\frac{1}{1\,+\,4G_V(M_Q/\Lambda_\chi)^2\Gamma(0,\,
(M_Q/\Lambda_\chi)^2)(1\,+\,\gamma_{01})}}\,,
\end{eqnarray}

\begin{eqnarray}               
\Lambda_\chi^2\,=\,\frac{2}{3}M_\rho^2G_V\Gamma(0,\,(M_Q/\Lambda_\chi)^2)(1\,
+\,\gamma_{03})\,,
\end{eqnarray}
\noindent
where $~\Gamma(n-2,\,(M_Q/\Lambda_\chi)^2)~$ is the incomplete gamma function 
\cite{BBdR}, and $~\gamma_{-1}~$, $~\gamma_{01}~$, and $~\gamma_{03}~$ are
perturbative corrections. Actually $~M_Q~$, $~\Lambda_\chi~$, and 
$~g_A~$ are taken as the input parameters to fit the data in Ref. \cite{BBdR}. 
In this paper, we calculate the contributions of the NJL Lagrangian
${\cal L}^{SP}_{NJL}+{\cal L}^{VA}_{NJL}$ to the precision of the leading
order in the $~1/N_c$-expansion, i.e. the uncertainty of the
calculation is of the order of $~30\%$. With this uncertianty, we shall
neglect the small quatities $~\gamma_{-1},~\gamma_{01}~{\rm and}~\gamma_{03}~$ 
throughout this paper. Moreover, we shall also neglect the non-factorized 
parts of the 4QC's given in Ref. \cite{WKYC} since they are of higher order in 
the expansion.

In calculating Feynman diagrams to the precision of the leading term in 
$1/N_c$-expansion, we should sum up all open-end chain diagrams attached to 
each quark line (cf. Fig.1). Note that only the flavor-singlet scalar four-
fermion interaction contributes to these chain diagrams. It is easy to see 
that the effect of summing up such chain diagrams attached to a quark line is 
just providing a dynamical mass term $~M_Q~$ (arising from $~\langle\bar{\psi}
\psi\rangle\neq 0~$) added to the current quark mass $~m~$ to form the full 
quark mass of this quark line. 
The relation 
\newpage
\null
\begin{center}
\epsfysize=2.5cm
\epsfig{file=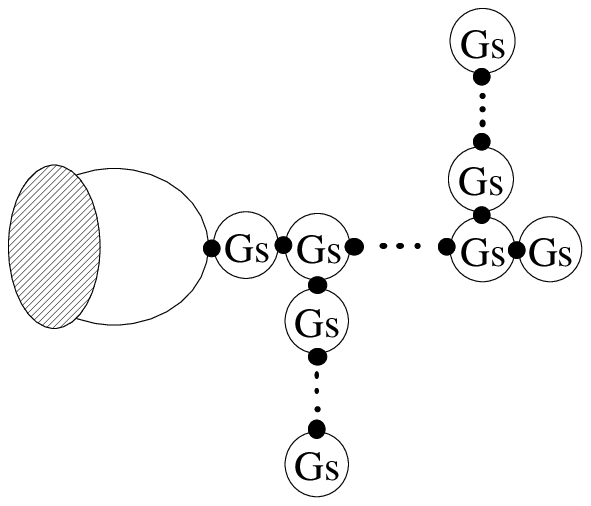}
\vspace{1.5cm}
\small{{\bf Fig.~1.}~Chain diagrams in the calculation of the leading 
contribution in $~1/N_c~$ expansion.}
\end{center}
\vspace{0.5cm}
between $~M_Q~$ and $~\langle\bar{\psi}\psi\rangle~$ can
be obtained as follows. In the large-$N_c~$ limit, the gap equation is 
\cite{BBdR}
\begin{eqnarray}                                            
\displaystyle \langle\bar{\psi}\psi\rangle=-\frac{3N_c}{4\pi^2}M_Q^3
\Gamma(-1,(M_Q^2/\Lambda_\chi^2))~,
\end{eqnarray}
in which the chiral limit relation $~\langle\bar{u}u\rangle=\langle\bar{d}d
\rangle=\langle\bar{s}s\rangle=\frac{1}{3}\langle\bar{\psi}\psi\rangle~$
is used. the function $~\Gamma(-1,(M_Q^2/\Lambda_\chi^2))~$ can be
eliminated by using (8). neglecting $~\gamma_{-1}~$ in (8), we get from
(11) and (8)
\begin{eqnarray}                                            
M_Q=\displaystyle -\frac{4\pi^2G_S}{3N_c\Lambda_\chi^2}
\langle\bar{\psi}\psi\rangle~.
\end{eqnarray}
Hence $~M_Q~$ is proportional to $~\langle\bar{\psi}\psi\rangle~$. Since
QCD sum rule concerns the VCE in which various dimensional quark condensates 
are put as explicit terms, all the effects of $~M_Q~$ should be put 
into the explicit quark condensates in the VCE by definition. Considering
these nonperturbative effects, (1) becomes
\begin{eqnarray}                                              
i\int d^4xe^{iq\cdot x}\langle
j^A(x)j^B(0)\rangle=F^{AB}(q)\sum_n\tilde{C}_n(Q^2)\langle O_n\rangle\,,
\end{eqnarray}
in which the new coefficient $~\tilde{C}_n(Q^2)~$ in the resulting VCE is not 
the same as the original Wilson coefficient $~C_n(Q^2)~$ in the standard OPE. 
$~\tilde{C}_n(Q^2)~$ contains nonperturbative effects from the NJL Lagrangian. 
Generally, $~\tilde{C}_n(Q^2)~$ is a function of $~\Lambda_\chi,~G_S,~G_V,~$ 
and the current quark mass $~m~$. {\it In the theory with the
Lagrangian {\rm (5)-(7)}, it is {\rm (13)} rather than {\rm (1)} that really 
matters in the QCD sum rule}. Note that {\it when calculating 
$~\tilde{C}_n(Q^2)~$, the quark mass in every quark line in the Feynman 
diagrams should be understood as the current quark mass 
$~m~$}. Since the current quark mass $~m~$ is much smaller than
$~Q^2\sim M^2_\rho~$, {\it we treat $~m^2/Q^2~$ as a small perturbation in
the following calculations.} \\

\begin{center}
{\bf III.~CALCULATION OF $~{\bf \tilde{C}_n}~$'s AND THE EXPLANATION OF
$~{\bf \kappa~}$'s}
\end{center}

In the sum rule for the $\rho$ meson, the currents $j^A,~j^B$ are taken
to be vector currents
\begin{eqnarray}                              
j_\mu=\displaystyle{ \frac{1}{2}(\bar{u}\gamma_\mu u-\bar{d}\gamma_\nu d)
= \bar{\psi}t_3\psi}\,,
\end{eqnarray}
where  $t_3$ is the flavor group generator in (6)-(7). The VCE
of the product of the currents takes the form
\begin{eqnarray}                               
&& \hspace{1.5cm}i\int dx~e^{iqx}~\langle 0|Tj_\mu(x)j_\nu(0)|0\rangle=(q_\mu 
q_\nu-q^2 g_{\mu\nu})\Pi(q^2)\,,\hspace{10cm}\\ \nonumber
&& \hspace{1.5cm}\Pi(Q^2)=\tilde{C}_I\langle I\rangle+\tilde{C}_{2Q}\langle
\bar{\psi}M_{u,d}\psi
\rangle+\tilde{C}_G\langle G^a_{\mu\nu}G^{a\mu\nu}\rangle
+\tilde{C}_{4Q}\sum_\Gamma\langle\bar{\psi}\Gamma\psi\bar{\psi}\Gamma\psi
\rangle+\cdots\,,\hspace{0.6cm}(15)
\end{eqnarray}
where $~M_{u,d}~$ is the $u-,~d-$quark mass matrix. We are 
going to calculate the four terms on the right-hand-side of (15) including 
the nonperturbative contributions from the NJL Lagrangian in (5)-(7).\\\\
\null\noindent
{\bf 1. Model with $~G_S\neq 0,~G_V=0$}.

There are five different ways of fitting the data presented in Ref. 
\cite{BBdR}, which determine different sets of values of $~M_Q~$, 
$~\Lambda_\chi~$, and $~g_A~$, and the predictions are all successful.
The simplest one of the fits is their {\bf Fit 4} in which $~g_A=1~$. From 
(9) we see that this corresponds to $~G_V=0~$ which makes the calculation of
the nonperturbative contributions easiest. Moreover, it is explained by
Weinberg \cite{W} that, in the large $N_c$ limit, $~g_A~$ should actually be 
unity in the constituent quark model. In view of this, we take the set 
{\bf Fit 4} in Ref. \cite{BBdR} in this subsection, and by means of (8)-(10), 
it leads to $~G_S=1.19,~G_V=0,~\Lambda_\chi=667~$ MeV. Since the uncertainty 
in our calculation is of the order of $30\%$, we are not taking these
\newpage
\begin{center}
\epsfysize=2.5cm
\epsfig{file=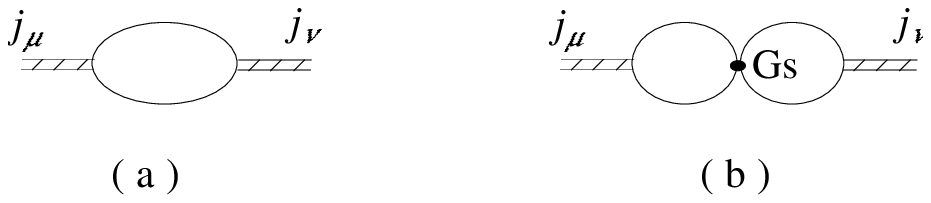}
\small{{\bf Fig.~2.}~Example of Feynman diagrams for leading contributions to\\
$~\tilde{C}_I\langle I\rangle~$.~(a)~PQCD contribution;~(b)~${\cal
L}^{SP}_{NJL}$ contribution.}
\vspace{0.5cm}

\end{center}
\vspace{0.5cm}
 precise values seriously and, instead, we shall consider a 
reasonable range of the parameters.

Now we calculate the coefficients in (15).
Examples of the Feynman diagram of the leading contribution of PQCD and 
${\cal L}^{SP}_{NJL}$ to $~\tilde{C}_I\langle I\rangle~$ is shown in Fig.2. 
This has been studied in Ref. \cite{YZ}. The first 
nonvanishing contribution comes from the term suppressed by $~G_S m^2/Q^2~$ 
which is {\it negligibly small} compared with our theoretical uncertainty. 
Thus {\it In the present appraoch $~\tilde{C}_I(Q^2)~$ is fairly 
well given by the standard Wilson coefficient $~C_I(Q^2)~$ from PQCD given 
in} Ref. \cite{SVZ}, i.e.
\begin{eqnarray}                                        
\tilde{C}_I\approx C_I=\displaystyle -\frac{1}{8\pi^2}
\ln\frac{Q^2}{\mu^2}\,.
\end{eqnarray}

The PQCD result of the $~\tilde{C}_{2Q}\langle\bar{\psi}M\psi\rangle~$ term is 
proportional to the current quark mass $~m~$ \cite{SVZ}~[cf. Fig.3(a)].
Examples of the Feynman diagram for the ${\cal L}^{SP}_{NJL}$ contribution to 
the $~\tilde{C}_{2Q}\langle\bar{\psi}M\psi\rangle~$ term is shown in Fig.3(b)
(This corresponds to the effect linear in $~M_Q~$.). 
It is easy to see that the first nonvanishing contribution is of the order of 
$~m\displaystyle\frac{m^2}{Q^2}~$ and is thus also {\it negligible} relative 
to the PQCD contribution to the present precision. Hence {\it $~\tilde{C}_{2Q}
(Q^2)~$ in the present approach is also mainly given by the standard Wilson 
coefficient $~C_{2Q}(Q^2)~$ from PQCD given in} Ref. \cite{SVZ}, i.e.
\begin{eqnarray}                                           
\tilde{C}_{2Q}\langle\bar{\psi}M_{u,d}\psi\rangle\approx C_{2Q}\langle
\bar{\psi}M_{u,d}\psi\rangle
=\displaystyle
\frac{1}{2Q^4}(m_u\langle\bar{u}u\rangle+m_d\langle\bar{d}d\rangle)\,.
\end{eqnarray}
\newpage
\begin{center}
\epsfysize=2.5cm
\epsfig{file=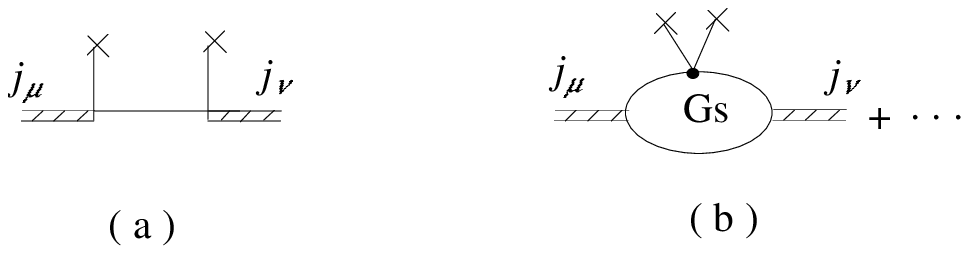}
\small{{\bf Fig.~3.}~Example of Feynman diagrams for leading contributions to 
$~\tilde{C}_{2Q}\langle\bar{\psi}M_{ud}\psi\rangle$.~(a)~PQCD contribution;~
(b)~${\cal L}^{SP}_{NJL}~$ contribution.}
\end{center}
\vspace{0.5cm}

The situation of the $~\tilde{C}_G\langle G^a_{\mu\nu}G^{a\mu\nu}\rangle~$ is 
similar. Examples of the Feynman diagrams for leading PQCD and 
${\cal L}^{SP}_{NJL}$ contributions to this term are shown in Fig.4. The first 
nonvanishing contribution in Fig.4(b) is also of order $~G_Sm^2/Q^2~$ which 
highly suppresses the nonperturbative contribution. Thus {\it 
$~\tilde{C}_G(Q^2)~$ is also maily given by the standard Wilson coefficient 
$~C_G(Q^2)~$ from PQCD 
given in }Ref. \cite{SVZ}, i.e.
\begin{eqnarray}                                           
\tilde{C}_G\langle G^a_{\mu\nu}G^{a\mu\nu}\rangle
\approx C_G\langle G^a_{\mu\nu}G^{a\mu\nu}\rangle
=\displaystyle \frac{\alpha_s}
{24\pi Q^4}\langle G^a_{\mu\nu}G^{a\mu\nu}\rangle~
\,.
\end{eqnarray}

For the 4QC term, the situation is completely different. Examples of the 
Feynman diagrams for leading order PQCD and ${\cal L}^{SP}_{NJL}$ 
contributions are shown in Fig.5. Unlike the previous cases (cf. Figs.2(b), 
3(b), 4(b), Fig.5(b) {\it is of tree-level rather than loop-level, therefore 
this contribution is large}. The leading PQCD [Fig.5(a)] formula given in Ref. 
\cite{SVZ} can be written as
\begin{eqnarray}                                                    
&& \hspace{1.5cm}\displaystyle{
\left.\sum_\Gamma \tilde{C}^\Gamma_{4Q}\langle\bar{\psi}\Gamma
\psi\bar{\psi}\Gamma\psi\rangle\right|_{PQCD}
=\sum_\Gamma C^\Gamma_{4Q}\langle\bar{\psi}\Gamma\psi\bar{\psi}\Gamma\psi
\rangle}\hspace{10cm}\\ \nonumber
&& \hspace{1.5cm}\displaystyle{
= -\frac{2\pi\alpha_s}{Q^6}} \left[\langle(\bar{\psi}\gamma_\mu\gamma_5 
\lambda^a t_3\psi)^2\rangle
\displaystyle{
+\frac{1}{9}}\langle\bar{\psi}\gamma_\mu
\lambda^a\left(\frac{1}{3}+\frac{1}{\sqrt 3}t_8\right)\psi
\bar{\psi}\gamma_\mu \lambda^a \psi\rangle\right]\,, \hspace{2.6cm}(19)
\end{eqnarray}
where $~C^\Gamma_{4Q}(Q^2)~$ is the standard Wilson coefficient. Note
that {\it there are only VEV's of products of vector and axial-vector 
currents}. The calculation of Fig.5(b) is straightforward but lengthy. There 
is also contribution to the 4QC term from the expansion of $~\langle
\bar{\psi}(x)\psi(0)\rangle~$ arising in Fig.3(b). Adding the two 
contributions together, we get 
\newpage
\begin{center}
\epsfysize=2.5cm
\epsfig{file=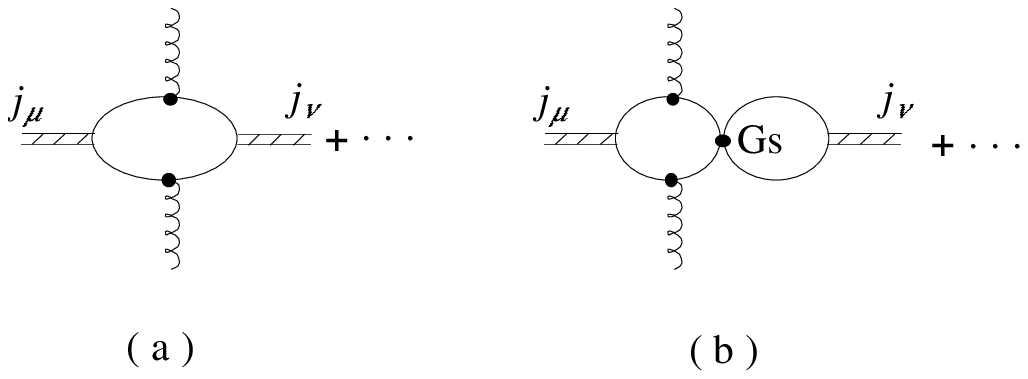}
\vspace{1.3cm}
\small{{\bf Fig.~4.}~Example of Feynman diagrams for leading contributions to
$~\tilde{C}_G\langle G^a_{\mu\nu}G^{a\mu\nu}\rangle$.~(a)~PQCD
contribution;~(b)~${\cal L}^{SP}_{NJL}~$ contribution.}
\end{center}
\vspace{0.5cm}
\begin{eqnarray}                                                      
&& \displaystyle 
\left.\sum_\Gamma \tilde{C}^\Gamma_{4Q} \langle\bar{\psi}\gamma\psi
\bar{\psi}\Gamma\psi\rangle\right|_{NJL}=-\frac{2\pi^2G_S}
{N_c\Lambda_\chi^2
Q^4}\left[\langle(\bar{\psi}\left(\frac{1}{3}+\frac{1}{\sqrt 3}+t_3
\right)\psi)^2\rangle
+\langle(\bar{\psi}
\left(\frac{1}{3}+\frac{1}{\sqrt 3}-t_3\right)\psi)^2\rangle\right.
\hspace{2.6cm}
\\ \nonumber
&& ~~~~\left.
-\langle\bar{\psi}t_+\psi\bar{\psi}
t_-\psi\rangle-\langle\bar{\psi}t_-\psi\bar{\psi}t_+
\psi\rangle-\langle(\bar{\psi}\gamma_5\left(\frac{1}{3}+\frac{1}{\sqrt
3}+t_3\right)\psi)^2\rangle
-\langle(\bar{\psi}\gamma_5\left(\frac{1}{3}+\frac{1}{\sqrt
3}-t_3\right)\psi)^2\rangle\right.
\hspace{1cm}\\ \nonumber
&& ~~~~\left.
+\langle\bar{\psi}\gamma_5 t_+\psi\bar{\psi}\gamma_5 t_-\psi
\rangle+\langle\bar{\psi}\gamma_5 t_-\psi\bar{\psi}\gamma_5 t_+\psi\rangle
\right]\,,\hspace{7.8cm}(20)\hspace{1cm}
\end{eqnarray}
where $~t_\pm\equiv t_1\pm it_2~$. We see that {\it {\rm (20)} contains the 
VEV's of products of scalar and pseudoscalar 4QC terms which are not suppressed 
by the current quark mass $~m~$, and thus are not small corrections to
{\rm (19)}}. Adding (19) and (20) together, we get the total 4QC term
\begin{eqnarray}                                                      
\displaystyle
\sum_\Gamma \tilde{C}^\Gamma_{4Q}\langle\bar{\psi}\Gamma\psi\bar{\psi}\Gamma
\psi\rangle =\left.\sum_\Gamma \tilde{C}^\Gamma_{4Q}\langle\bar{\psi}\Gamma
\psi\bar{\psi}\Gamma\psi\rangle\right|_{PQCD}
+\left.\sum_\Gamma \tilde{C}^\Gamma_{4Q}\langle\bar{\psi}\Gamma\psi\bar{\psi}
\Gamma\psi\rangle\right|_{NJL}\,.
\end{eqnarray}
{\it Compared with the standard OPE formula {\rm (19)} used in Ref. 
{\rm\cite{LNT}, (21)} contains extra significant terms which enhance
the 4QC term in the VCE {\rm (15)}}.

The above results are consistent with the analysis in Ref. \cite{LNT}, i.e.
the first three terms in (15) are as usual, while the 4QC term is
enhanced. With (21), we can define the enhancement factor $~\kappa_{4Q}~$ in
this approach as 
\begin{eqnarray}                                          
\displaystyle
\kappa_{4Q}=\frac{\displaystyle \sum_\Gamma \tilde{C}^\Gamma_{4Q}\langle
\bar{\psi}\Gamma\psi\bar{\psi}\Gamma\psi\rangle}{\displaystyle \sum_\Gamma 
C^\Gamma_{4Q}\langle\bar{\psi}\Gamma\psi\bar{\psi}\Gamma\psi\rangle}\,,
\end{eqnarray}
where the numerator is given by (21) and the denominator is given by (19).
The 4QC's in (19) and (21) contain their factorized parts given by (2)
and non-factorized parts given in 
\newpage
\begin{center}
\epsfysize=2.5cm
\epsfig{file=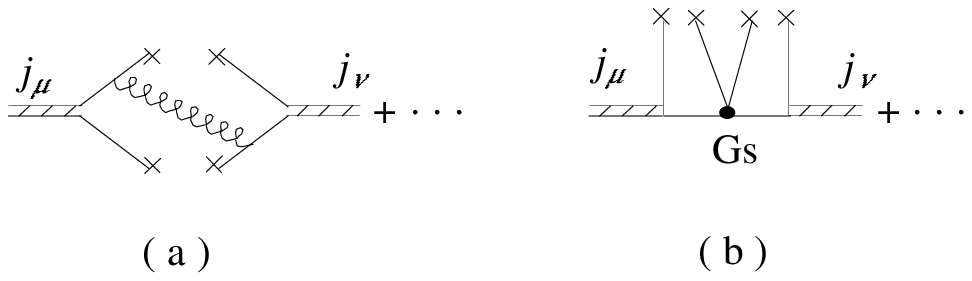}
\vspace{1cm}
\small{{\bf Fig~.5}~Example of Feynman diagrams for leading contributions to
 $\sum_\Gamma \tilde{C}^\Gamma_{4Q}\langle\bar{\psi}\Gamma\psi\bar{\psi}\Gamma
\psi\rangle$ .~(a)~PQCD contribution;~(b)~${\cal L}^{SP}_{NJL}~$contribution.}
\end{center}
\vspace{0.5cm}
Ref.\cite{WKYC}. As we have stated,
to the present precision we neglect the non-factorized parts. The
factorized part of (19) reads \cite{LNT}
\begin{eqnarray}                                           
\displaystyle
\sum_\Gamma\tilde{C}^\Gamma_{4Q}
\langle\bar{\psi}\Gamma\psi\bar{\psi}
\Gamma\psi\rangle\bigg|_{^{PQCD}_{factorized}} =-\frac{14\pi\alpha_s}{81}
\left(1-\frac{1}{N_c^2}\right)\langle\bar{\psi}\psi\rangle^2\,.
\end{eqnarray}
The factorized part of (20) obtained from (2) is
\begin{eqnarray}                                           
\displaystyle
\sum_\Gamma\tilde{C}^\Gamma_{4Q}\langle\bar{\psi}\Gamma\psi\bar{\psi}
\Gamma\psi\rangle\bigg|_{^{NJL}_{factorized}}=-\frac{4\pi^2G_S}{9N_c
\Lambda_\chi^2 Q^4}\langle\bar{\psi}\psi\rangle^2\,.
\end{eqnarray}
From (12) we see that these factorized parts correspond to $~M_Q^2~$
contributions to the 4QC term. The explicit 
formula for $~\kappa_{4Q}~$ for a given $~Q^2~$ in the present approach is 
\begin{eqnarray}                                              
\displaystyle
\kappa_{4Q}\approx 1+\frac{18\pi N_cG_S Q^2}{7(N_c^2-1)\alpha_s\Lambda_\chi^2}
\,.
\end{eqnarray}
We now estimate the value of $~\kappa_{4Q}~$ in (25). We take $~N_c=3~$ and 
the PQCD coupling constant $~\alpha_s(M_\rho)\approx 0.48~$ which is obtained 
from the two-loop evolution formula from the world average value 
$~\alpha_s(M_Z)=0.118~$ \cite{PDG}. From (8) we see 
that $~G_S~$ and $~M_Q^2/\Lambda^2_\chi~$ are related. We take $~M_Q=199~$
MeV from {\bf Fit 4} in Ref. \cite{BBdR}, and take $~\Lambda_\chi~$ around 
the chiral symmetry breaking scale $~\Lambda_\chi\sim 4\pi f_\pi\sim 1.2~$GeV. 
The values of $~\kappa_{4Q}~$ with $~Q^2=M^2_\rho~$ for various values of 
$~\Lambda_\chi~$ are shown in Table 1.
\newpage
\vspace{0.2cm}
\begin{center}
\small{{\bf Table 1}~ Values of $~\kappa_{4Q}~$ with $~Q^2=M^2_\rho~$ for 
various given values of $~\Lambda_\chi~$ in the $~G_S\neq 0,~G_V=0~$ model.}\\
\vspace{0.5cm}
\normalsize
\tabcolsep 5pt
\begin{tabular}{*{3}{c}}
\hline
\hline
\multicolumn{1} {l}{$\;\;\Lambda_\chi$~(GeV)} & {$\;\;\;\;\;\;\;\;\;\;G_S$}
 & {$\;\;\;\;\;\;\;\;\;\;\kappa_{4Q}\;\;$}\\
\hline
{$\;\;1.25$} & {$\;\;\;\;\;\;\;\;\;\;1.10$} & {$\;\;\;\;\;\;\;\;\;\;3.7\;\;$}\\
{$\;\;1.10$} & {$\;\;\;\;\;\;\;\;\;\;1.15$} & {$\;\;\;\;\;\;\;\;\;\;4.7\;\;$}\\
{$\;\;0.90$} & {$\;\;\;\;\;\;\;\;\;\;1.20$} & {$\;\;\;\;\;\;\;\;\;\;6.7\;\;$}\\
{$\;\;0.80$} & {$\;\;\;\;\;\;\;\;\;\;1.25$} & {$\;\;\;\;\;\;\;\;\;\;8.3\;\;$}\\
\hline
\hline
\end{tabular}
\end{center}
\vspace{0.4cm}
\noindent
We see that, with the uncertainty of the present calculation, {\it the values
of $~\kappa_{4Q}~$ are consistent with the values in {\rm (3)}.}
Actually in the analysis in Ref. \cite{LNT}, Borel transformation
\begin{eqnarray}                                                
\displaystyle
\lim_{^{Q^2\to\infty,~n\to\infty}_{~Q^2/n\to M^2}}\frac{1}{(n-1)!}(Q^2)^n
(-\frac{d}{dQ^2})^n
\end{eqnarray}
was applied. We have also looked at the result after applying the Borel
transformation (26) with $~M\approx M_\rho~$. The results of $~\kappa_{4Q}~$ 
is larger than those listed in Table 1. For example, for $~\Lambda_\chi=1.25~$
GeV and $1.10$ GeV, we have $~\kappa_{4Q}=6.5~$ and $8.3$,
respectively. {\it These are still consistent with the phenomenological value 
in {\rm (3)} considering the uncertainty of the present calculation}. 
Thus {\it the present theory with $~G_S\neq 0~{\rm and}~G_V=0~$ is compatible 
with the analysis in Ref. {\rm \cite{LNT}} that the first three terms in the 
VCE {\rm (15)} are the same as those based on the basic assumption {\rm (i)}, 
while the 4QC term should be enhanced by a factor $~\kappa_{4Q}~$ given
in {\rm (3)}}.\\\\
\null\noindent
{\bf 2. Model with $~G_S\neq 0,~G_V\neq 0$.}

 The above model with $~G_S\neq 0,~G_V=0~$ is not compatible with the 
enhancement  factors (4) given in Ref. \cite{BDLPR} in which
$~\kappa_G~$ is not unity. Since the ${\cal L}^{VA}_{NJL}$ interactions give 
significant contributions to the first three terms in (15), it is interesting 
to see if a model with $~G_S\neq 0~{\rm and}~G_V\neq 0~$ can be compatible 
with (4).

We first consider the ${\cal L}^{VA}_{NJL}$ contributions to the 4QC term.
The Feynman diagam for the leading contribution is the same as Fig.5(b) with
${\cal L}^{SP}_{NJL}$ replaced by ${\cal L}^{VA}_{NJL}$. After
straightforward calculations, we find that the factorized
parts of $~\langle(\bar{\psi}\gamma_\mu t_3\psi)^2\rangle~$ and
$~\langle(\bar{\psi}\gamma_\mu\gamma_5 t_3\psi)^2\rangle~$ cancel
each other so that the net contribution of the ${\cal L}^{VA}_{NJL}$ 
interaction to the 4QC term is just the corrections from the
non-factorized parts which is {\it negligible}. So that {\it the value of 
$~\kappa_{4Q}~$ is not significantly affected by $~G_V\neq 0~$}. Let us take 
{\bf Fit 1} in Ref. \cite{BBdR} as an example in which
$~g_A=0.61~$,$~\Lambda_\chi=1.16~{\rm GeV},~M_Q=265~{\rm MeV}~$ 
($\gamma_{-1}=\gamma_{01}=\gamma_{03}=0~$ in 
{\bf Fit 1}). Fixing this $M_Q$, the values of $~G_S~$ and 
$~\kappa_{4Q}~$ with $~Q^2=M^2_\rho~$ for various values of $~\Lambda_\chi~$ 
are listed in Table 2. We see that {\it these values are consistent with 
{\rm (4)} considering the uncertainty of the present approach.}

\vspace{0.2cm}
\begin{center}
\small{{\bf Table 2}~ Values of $~\kappa_{4Q}~$ and $~\kappa_G~$ with
$~Q^2=M^2_\rho~$ for various given values of $~\Lambda_\chi~$ in the 
$~G_S\neq 0,~G_V\neq 0~$ model.}\\
\vspace{0.5cm}
\normalsize
\tabcolsep 5pt
\begin{tabular}{*{5}{c}}
\hline
\hline
\multicolumn{1} {l}{$\;\;\Lambda_\chi$~(GeV)} & {$\;\;\;\;\;\;\;\;\;\;G_S$}
 & {$\;\;\;\;\;\;\;\;\;\;G_V$} & {$\;\;\;\;\;\;\;\;\;\;\kappa_{4Q}$}
 & {$\;\;\;\;\;\;\;\;\;\;\kappa_G \;\;$}\\
\hline
{$\;\;1.16$} & {$\;\;\;\;\;\;\;\;\;\;1.22$} & {$\;\;\;\;\;\;\;\;\;\;1.26$}
 & {$\;\;\;\;\;\;\;\;\;\;4.4$} & {$\;\;\;\;\;\;\;\;\;\;2.5\;\;$}\\
{$\;\;1.00$} & {$\;\;\;\;\;\;\;\;\;\;1.28$} & {$\;\;\;\;\;\;\;\;\;\;1.06$}
 & {$\;\;\;\;\;\;\;\;\;\;5.8$} & {$\;\;\;\;\;\;\;\;\;\;2.7\;\;$}\\
{$\;\;0.90$} & {$\;\;\;\;\;\;\;\;\;\;1.40$} & {$\;\;\;\;\;\;\;\;\;\;0.94$}
 & {$\;\;\;\;\;\;\;\;\;\;7.5$} & {$\;\;\;\;\;\;\;\;\;\;2.9\;\;$}\\
{$\;\;0.80$} & {$\;\;\;\;\;\;\;\;\;\;1.42$} & {$\;\;\;\;\;\;\;\;\;\;0.84$}
 & {$\;\;\;\;\;\;\;\;\;\;9.3$} & {$\;\;\;\;\;\;\;\;\;\;3.1\;\;$}\\
\hline
\hline
\end{tabular}
\end{center}
\vspace{0.4cm}
\noindent

We then look at the ${\cal L}^{VA}_{NJL}$
contributions to the gluon condensate term (the two-quark condensate
term is relatively much smaller). The Feynman diagram is still Fig.4(b) with
${\cal L}^{SP}_{NJL}$ replaced by ${\cal L}^{VA}_{NJL}$. It is easy to
see that only the $~(\bar{\psi}\gamma_\mu t_3\psi)^2~$ term in (7) contributes.
This gives rise to 
\begin{eqnarray}                                             
\displaystyle
\tilde{C}_G(Q^2)=\left.\tilde{C}_G(Q^2)\right|_{PQCD}+\left.\tilde{C}_G(Q^2)
\right|_{NJL}\approx\left[1+\frac{4G_VQ^2}{N_c\Lambda_\chi^2}~\ln\frac{Q^2}
{\mu^2}\right]C_G(Q^2)\,.
\end{eqnarray}
Thus we have the enhancement factor
\begin{eqnarray}                                             
~\displaystyle \kappa_G\equiv \frac{\tilde{C}_G(Q^2)}{C_G(Q^2)}
\approx {1+\frac{4G_VQ^2}{N_c\Lambda_\chi^2}~\ln\frac{Q^2}{\mu^2}}~.
\end{eqnarray}
In the QCD sum rule, the scale $~\mu~$ is supposed to be the inverse of
the confinement radius, $~R^{-1}_{conf}~$ \cite{SVZ}. Thus 
$~\mu\approx \Lambda_{\overline{MS}}\approx 280~$MeV. The values of
$~G_V~$ and $~\kappa_G~$ with $~Q^2=M^2_\rho~$ for various values of 
$~\Lambda_\chi~$ are listed in Table 2. We see that {\it these are also 
consistent with the $~\kappa_G~$ in {\rm (4)}}. 

The ${\cal L}^{VA}_{NJL}$ interactions will also give rise to the same
enhancement factor as $~\kappa_G~$ for the 
$~\tilde{C}_I \langle I\rangle~$ term \cite{YZ}. For $~Q^2\approx M^2_\rho>
\mu^2~$, the PQCD value of $~C_I~$ [cf. (16)] is negative. Therefore this 
nonperturbative contribution makes $~\tilde{C}_I~$ more negative. We expect 
a global analysis in the spirit of Ref. \cite{BDLPR} including the 
$~\tilde{C}_I\langle I \rangle~$ term and test the present prediction.\\

\begin{center}
{\bf IV.~CONCLUSIONS}
\end{center}

In this paper, taking the $~\rho$-meson sum rule as an example, we have tried 
to modify the basic assumption of the QCD sum rule that the 
coefficients $~C_n(Q^2)~$'s in (1) are merely determined by PQCD, and studied 
the possibility of explaining the phenomenological enhancement factors for
certain terms in (1) \cite{LNT}\cite{BDLPR} by taking account of 
nonperturbative contributions to the VCE which lead to (15). We take the QCD 
motivated ENJL model in Ref. \cite{BBdR} as the low energy effective 
Lagrangian for QCD, and have calculated the nonperturbative 
contributions from the NJL Lagrangian to obtain the new coefficients 
$~\tilde{C}_n~$'s to the precision of the leading order in the $~1/N_c~$ 
expansion.

It is interesting that taking {\bf Fit 4} of Ref. \cite{BBdR}, ($~G_S\neq 0,~
G_V=0$), our result shows that {\it the first three terms in {\rm (15)} are not
affected much by the nonperturbative interactions, while the 4QC term
contains extra terms with new tensor structures coming from the 
nonperturbative interactions and they are large compared with the PQCD
result}. Regarding these extra contributions as the theoretical
source of the enhancement factor $~\kappa_{4Q}~$, {\it the obtained 
$~\kappa_{4Q}~$} ({\it cf. {\rm Table 1)} is consistent with the 
phenomenological value {\rm (3)}, considering the theoretical uncertainty in 
the present calculation, for reasonable range of the parameters in the ENJL 
model}. 

Taking {\bf Fit 1} of Ref. \cite{BBdR}, i.e. $~G_S\neq 0,~G_V\neq 0$, we
have got another result that the obtained $~\kappa_{4Q}~$ and $~\kappa_G~$ 
listed in Table 2  {\it are all consistent with the results {\rm (4)} given in 
Ref. {\rm \cite{BDLPR}}}. However, in this model, $~\tilde{C}_I\langle I
\rangle~$ will also be larger than $~C_I\langle I\rangle~$ 
by the same factor as $~\kappa_G~$ listed in Table 2, but this has not been 
discussed in Ref. \cite{BDLPR}. 
We expect a further analysis of the $~\tilde{C}_I\langle I\rangle~$ term in 
the spirit of Ref. \cite{BDLPR} to test whether the present prediction by 
{\bf Fit 1} is really right.

We conclude that {\it it is possible to explain the enhancement
factors in both Ref. {\rm \cite{LNT}} and {\rm \cite{BDLPR}} by taking 
different sets of parameters in the ENJL model in Ref. {\rm \cite{BBdR}}}. 
Within the $~30\%~$ uncertainty of the present calculation, the model with 
$~G_S\neq 0,~G_V=0~$ can give a reasonable explanation of the analysis in
Ref. \cite{LNT} with the value of $~\kappa_{4Q}~$ given in (3), and the
model with $~G_S\neq 0,~G_V\neq 0~$ ENJL model \cite{BBdR} can possibly 
explain the enhancement factors (4) given in Ref. \cite{BDLPR}. In the 
present approach, {\it the physical origin of the enhancement factors is 
essentially the nonperturbative contributions to the VCE which makes the
VCE different from the original assumption {\rm (i)}}, whereas the non-
factorized part of the 4QC given in Ref. \cite{WKYC} is only a minor source 
of the enhancement factors. More precise nonperturbative
QCD approach along this line is expected for developing practically useful 
algorithm to improve the application of the QCD sum rule to low energy 
processes.

\begin{center}
{\bf Acknowledgement}
\end{center}

This work is supported by the National Natural Science Foundation of China,
the Fundamental Research Foundation of Tsinghua University, and a
special grant from the State Commission of Education of China. We would like 
to thank K. Yamawaki and R.A. Bertlmann for interesting discussions.


\end{document}